\documentclass[prl,twocolumn,amssymb,superscriptaddress]{revtex4-2}
\usepackage{hyperref,amsmath,graphicx,subfigure}
\usepackage{geometry}
\newcommand{\arXiv}[1]{\href{http://www.arXiv.org/abs/#1}{arXiv:#1}}
\renewcommand{\doi}[2]{\href{http://dx.doi.org/#2}{#1}}
\newcommand{\beq}{\begin{equation}}
\newcommand{\eeq}{\end{equation}}
\DeclareMathOperator{\Tr}{Tr}

\begin{document}

\title{Unified theory of local integrals of motion}

\author{Ben Craps} 
\affiliation{Theoretische Natuurkunde, Vrije Universiteit Brussel (VUB) and International Solvay Institutes, 1050 Brussels, Belgium}
\author{Oleg Evnin}
\affiliation{\mbox{High Energy Physics Research Unit, Faculty of Science, Chulalongkorn University, Bangkok 10330, Thailand}}
\affiliation{Theoretische Natuurkunde, Vrije Universiteit Brussel (VUB) and International Solvay Institutes, 1050 Brussels, Belgium}
\author{Dmitry Kovrizhin}
\affiliation{Laboratoire de Physique Th\'eorique et Mod\'elisation (LPTM), CY Cergy Paris Universit\'e, CNRS, F-95302 Cergy-Pontoise, France}
\author{Gabriele Pascuzzi}
\affiliation{Theoretische Natuurkunde, Vrije Universiteit Brussel (VUB) and International Solvay Institutes, 1050 Brussels, Belgium}

\begin{abstract}
Conservation laws are of paramount importance in our understanding of classical and quantum dynamics. Here, we present a general framework for constructing exact quantum integrals of motion with the desired locality and quantum numbers, which will be illustrated for the case of many-body-localization (MBL). The latter has been understood theoretically in terms of the existence of an extensive number of local integrals of motion (LIOMs). Using our approach, we show that one can express the task of finding LIOMs as an optimization problem. For some specifications, this problem surprisingly connects to the question of finding classical ground states of spin-glass models. Our work unifies previous results obtained in the MBL context and reveals intriguing connections between many-body localization, spin-glass physics, and constrained optimization problems.
\end{abstract}

\maketitle

Conservation laws, or the `integrals of motion,' have played a defining role in our understanding of dynamics, including the seminal developments in the theory of integrability as well as Noether's relation between conservation laws and symmetries. In quantum theories, conservation laws are abundant, as any projector on Hamiltonian eigenstates is a conserved operator, but a crucial question is whether conservation laws exist that are simple (local, few-body) expressions in terms of the underlying local degrees of freedom. Our aim here is to outline a systematic procedure for recovering such conservation laws and to demonstrate how it operates in practice. The approach we present is very general and can be used in any quantum system that possesses local conservation laws. In our illustrations below, we will focus on disordered spin chains for concreteness.

One arena where the topic of quantum conservation laws has recently attracted a lot of attention \cite{MBL1,MBL2,MBL3,review} is many-body localization (MBL). Static disorder in the presence of interactions has been shown to suppress transport in a way that parallels Anderson localization in noninteracting systems \cite{anderson}. This discovery has generated extensive research activity on non-equilibrium dynamics of disordered quantum matter, including recent realizations in experiments with cold atoms \cite{Bloch, Greiner}, and programmable quantum simulators \cite{Shtanko, Roushan}. 

A central theoretical insight is that the MBL phase admits an extensive set of local integrals of motion~\cite{SPA,RosMuellerScardicchio}.  By definition, a LIOM commutes with the Hamiltonian and is predominantly supported on a few neighboring sites.  However, the choice of these conserved quantities is not unique, as any linear combination of LIOMs is again an integral of motion.  A systematic framework that removes this ambiguity and makes the locality requirements explicit is therefore desirable.

Several construction schemes for LIOMs have been put forward previously.  In the approach of~\cite{CHVA}, the time evolution of a single spin is averaged over a fixed time interval, converging to an exactly conserved quantity in the long-time limit.  The authors of~\cite{RO} apply a sequence of unitary transformations to a single-spin operator, where an exact integral of motion emerges after infinitely many steps.  Approximate LIOMs with strictly bounded support are built in~\cite{OAVP}. In some of these approaches, such as~\cite{RO,PLYWC,AAS},  constraints on the quantum numbers of the LIOMs are enforced, while in others~\cite{CHVA,OAVP}, they remain unconstrained. (Further recent considerations of LIOMs can be seen in \cite{recent1,recent2,recent3}.)

A common feature of these schemes is that locality is not imposed {\it a priori}, but instead arises as an emergent property when the Hamiltonian exhibits MBL. In other words, one starts from a simple operator (e.g., a single spin) and the procedure yields an exactly conserved quantity whose spatial profile becomes localized as long as the Hamiltonian is in the MBL phase. By contrast, the present work introduces a formalism in which the desired form of locality is imposed at the start of the calculations. By specifying, for example, the maximal support (one site, two adjacent sites, etc.), our approach returns an exact integral of motion that is optimal with respect to the required locality criteria. This construction unifies and extends previous methods by providing a controlled way to generate LIOMs with a prescribed spatial range and quantum numbers. It will also demonstrate surprising connections between constructing the `pseudospin' local conservation laws with eigenvalues $\pm 1$ and the physics of spin glasses \cite{SKM} and specifically the structure of the Hopfield model \cite{HopN,HopSG}, leading to a relation between quantum many-body physics and these other actively investigated domains of research.

{\it Q-matrix formalism}. Our approach to constructing LIOMs relies on a general procedure for building conservation laws that preferentially select certain directions in the operator space. This procedure, known as the Q-matrix formalism, first arose in works on Nielsen complexity of quantum evolution~\cite{bound,complint}. Consider a quantum system with a finite-dimensional Hilbert space of dimension $D$. The system is described by a Hamiltonian $\hat{H}$ whose eigenstates are labeled $|n\rangle$, with the corresponding eigenvalues $E_n$. In the following, we assume that the spectrum of $\hat{H}$ is non-degenerate (and refer to \cite{complint} for ways to relax this assumption). Our goal is to find a conserved operator that maximizes a certain objective function. To this end, we introduce a complete basis $\{\hat{T}_a\}$ of Hermitian operators orthonormal with respect to the Hilbert-Schmidt inner product $\mathrm{Tr}(\hat{T}_a\hat{T}_b)=\delta_{ab}$. Each basis element is assigned a weight $w_a\in[0,1]$ that is chosen to reflect a specific locality profile -- for example, $w_a=1$ for all basis operators acting exclusively on a specific site, and $w_a=0$ for all basis operators with support away from this site.
For any Hermitian operator $\hat{A}=\sum_a c_a \hat{T}_a$, we define the objective function $R$ to be maximized later,
\beq\label{eq_R}
R[\hat{A}]\equiv \sum_a w_a c^2_a[\hat{A}],\ \ \ \ c_a[\hat{A}]\equiv\Tr(\hat{A}\hat{T}_a),
\eeq
where the choice of a quadratic dependence on $c_a$ is motivated by the relative simplicity of the optimization problems that will arise from this definition.
In the following, we will consider operators $\hat V$  that are integrals of motion of $\hat H$ so that $[\hat{V},\hat{H}]=0$. Any such operator with eigenvalues $v_n$ can be written in the eigenbasis of $\hat{H}$ as 
\beq\label{Hvn}
\hat{V}=\sum_n v_n\,|n\rangle\langle n|.
\eeq
Its expansion coefficients in the basis of operators $\{\hat{T}_a\}$ are given by $c_a[\hat{V}]=\sum_{n}v_n\langle n|\hat{T}_a|n\rangle$, and substituting these into the definition of the objective function (\ref{eq_R}) yields a quadratic form
\beq\label{Rv}
R[\hat{V}]=\sum_{mn} Q_{mn}v_m v_n,
\eeq
where the $Q$-matrix is defined as
\beq
Q_{mn}\equiv\sum_a w_a\langle n|\hat{T}_a |n\rangle\langle m|\hat{T}_a |m\rangle.
\eeq
This matrix of dimension $D$ is real, symmetric, and positive semi-definite. A requirement that $R[\hat{V}]$ should be maximized under the normalization constraint $\sum_n v_n^{2}=1$ leads to a standard eigenvalue problem, where the optimal vector $\{v_n\}$ is the eigenvector of $Q$ associated with its largest eigenvalue. For the case of LIOMs, we will choose the weights $w_a$ so that these eigenvectors correspond to observables with support on a single site or on a few adjacent sites. Additional constraints, such as orthogonality to previously obtained LIOMs, can be incorporated within the same variational framework.

\textit{Construction of $l$-bits.} Fixing only the overall normalization does not allow one to control the quantum numbers $v_n$ corresponding to $\hat{V}$. In many constructions (e.g. Refs.~\cite{RO,PLYWC,AAS}), one wishes the LIOMs to have the same eigenvalue spectrum as the microscopic degrees of freedom. For a spin-$\tfrac12$ system, we therefore require $v_n\in\{+1,-1\}$ for every $n$. Operators with such a binary spectrum may be referred to as `localized bits' or {\it $l$-bits} \cite{MBL3}. The vector $\mathbf v$ is then an Ising spin configuration with zero net magnetization (the latter follows from the trace condition $\sum_n v_n=0$).  Maximization of $R[\hat{V}]$ with this binary constraint is related to solving the Quadratic Unconstrained Binary Optimization (QUBO) problem
$\max_{\mathbf v\in\{\pm1\}^{D}} \mathbf v^{\!\top} Q\,\mathbf v$, which is in the NP-hard class, see e.g.~\cite{Kochenberger, Pardalos, Anderson_Ising}.

In practice, we treat the binary case as a problem of finding the ground-state of an Ising spin-glass, which is hard in general. However, when the assigned weights favor only $z$-components of spins that are restricted to a few sites, the low-rank structure of $Q$ permits an efficient numerical implementation. Our construction can be applied blockwise if $\hat H$ possesses symmetry sectors (e.g.,~fixed total $\hat{S}^z$). In this case, one restricts indices $n$ to a given sector and repeats the procedure, thereby preserving all conserved quantum numbers. The set of operators $\hat V^{(\alpha)}$ obtained from the eigenvectors of $Q$ plays precisely the role of the $l$-bits $\hat\tau_i$ which `diagonalize' the Hamiltonian. Thus, our variational scheme provides a systematic way to construct the $l$-bit basis directly from the microscopic Hamiltonian.

\textit{Local integrals of motion in an MBL model of spins}. In what follows, we illustrate the above procedure in several concrete setups, both with and without eigenvalue constraints. Our primary example is the one-dimensional random-field Heisenberg chain with periodic boundary conditions, which has been extensively studied previously in considerations of MBL physics:
\beq\label{chain}
\hat{H}=\sum_{i=1}^{L}(\hat{S}_i^{x}\hat{S}_{i+1}^{x}+\hat{S}_i^{y}\hat{S}_{i+1}^{y}+\hat{S}_i^{z}\hat{S}_{i+1}^{z})+\sum_{i=1}^{L} h_i \hat{S}_i^{z},
\eeq
where random magnetic fields $h_i$ are drawn uniformly in $[-W,W]$. In the previous numerical calculations, it was observed that this Hamiltonian displays an MBL transition which occurs at $W=3.5\pm1.0$ \cite{transition}. We have used the QuSpin package \cite{quspin} for the numerical simulations of spin chains. For pedagogical purposes, in the supplemental material, we also apply our exact construction to the single-particle case of Anderson localization.

\textit{Maximization of the overlap with a single spin}. The central feature of the MBL phase is that there exist integrals of motion having large overlaps with single spin operators $\hat{S}^z_i$, while the overlaps with other spins fall off exponentially with the distance. In order to search for these LIOMs, we chose the set of operators $\hat{T}_a$ given by a complete Pauli basis on the whole spin-chain, and we set a single $w_a=1$ corresponding to the operator $\hat{S}^z_j$ for a specific site $j$, while $w_a=0$ for all other operators. For these choices, the matrix $Q$ assumes a simple form:
\beq\label{Qsing}
Q^{(j)}_{mn}=\langle n|\hat{S}_j^z|n\rangle\langle m|\hat{S}_j^z|m\rangle.
\eeq
The solution of the maximization problem of \eqref{Rv} together with \eqref{Qsing} is the top eigenvector of $Q_{mn}^{(j)}$, 
\beq\label{vj}
v^{(j)}_n = \langle n|\hat{S}^z_{j}|n\rangle,
\eeq
and the corresponding conserved operator is given by $\hat V^{(j)}=\sum_{n}v^{(j)}_n|n\rangle\langle n|$. This result can be compared with the one obtained via infinite-time averaging of the spin operator $\hat{S}^z_j(t)=\sum_{mn}e^{i(E_m-E_n)t} \langle m| \hat{S}^z_j |n\rangle |m\rangle\langle n|$,
\beq\label{sz_avg}
\overline{\hat{S}^z_j} = \lim_{T\rightarrow\infty} \frac{1}{T} \int_0^\infty \hat{S}^z_j(t)dt = \sum_n \langle n|\hat{S}^z_j|n\rangle |n\rangle\langle n|.
\eeq
This shows that maximization of the overlap with a single onsite operator gives the same LIOMs as the ones derived from the time-averaging prescription in \cite{CHVA}.
Moreover, conservation of $\hat{S}^z_\text{tot}$ causes the diagonal matrix elements of $\hat{S}^x_j$ and $\hat{S}^y_j$ in the eigenbasis to vanish. Therefore, $Q$ is unchanged when introducing weight 1 for these two operators, which means that for this system, \eqref{sz_avg} is in fact the maximally localized operator on site $j$.

\begin{figure}[t]
\centering
\mbox{\subfigure{\includegraphics[width=0.23\textwidth]{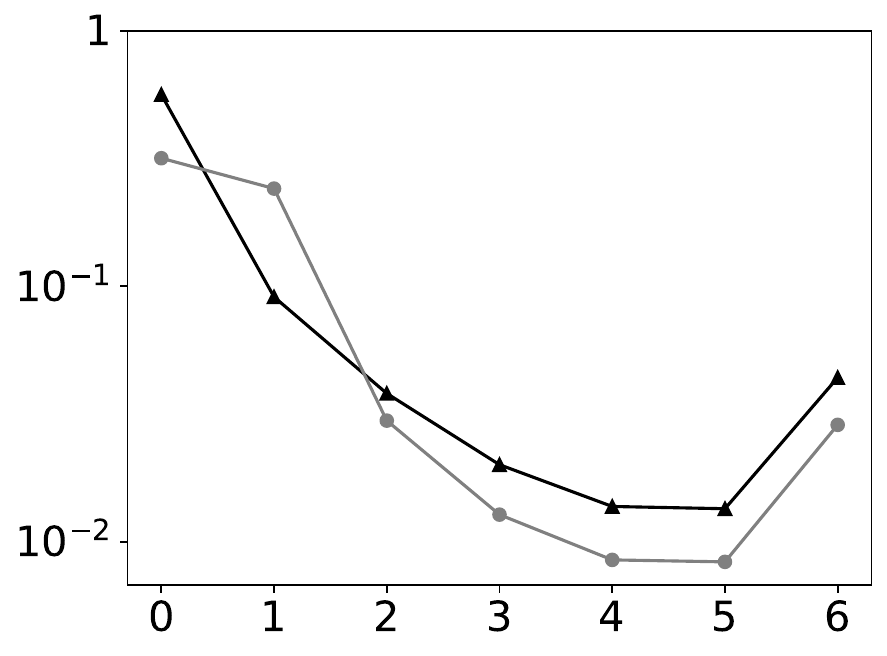}}\,
 \subfigure{\includegraphics[width=0.23\textwidth]{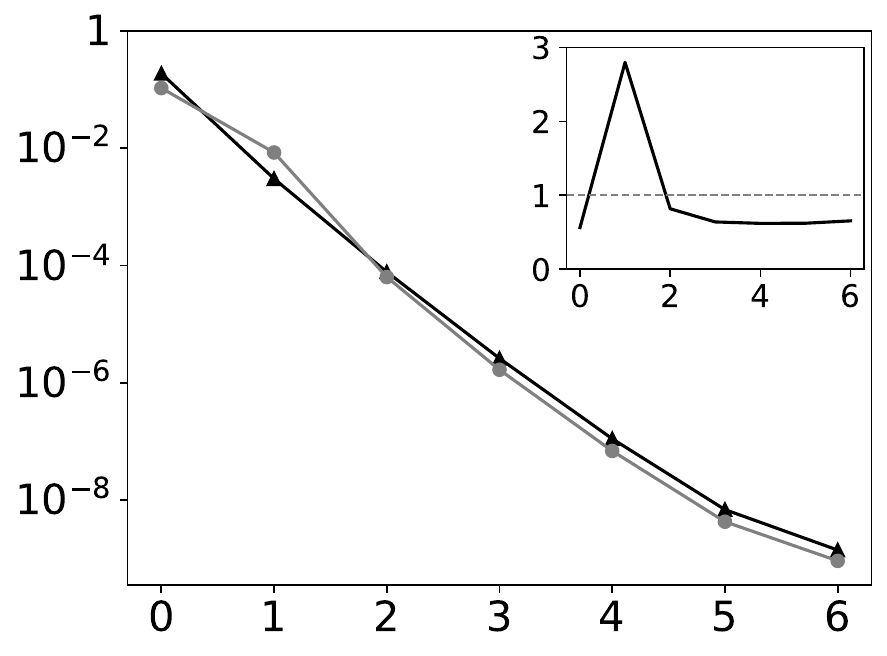}}}
\begin{picture}(0,0)
\put(-140,-5){$d$}
\put(-237,73){$p_i$}
\put(-20,-5){$d$}
\put(-117,73){$p_d$}
\put(-62,73){$\mathrm{\scriptstyle ratio}$}
\put(-16,43){$\scriptstyle d$}
\end{picture}
\caption{Support probabilities for the LIOMs maximized on a single site (black triangles), and on three adjacent sites (grey circles). Results are averaged over 1000 disorder realizations. {\it Left panel:} total probability for all operators at a given distance $d$ from the central site. {\it Right panel:} average probability per operator. {\it Inset:} ratio of $p_d$ for the LIOMs optimized on three sites to $p_d$ for the single site optimization.}
\label{fig:od}
\end{figure}
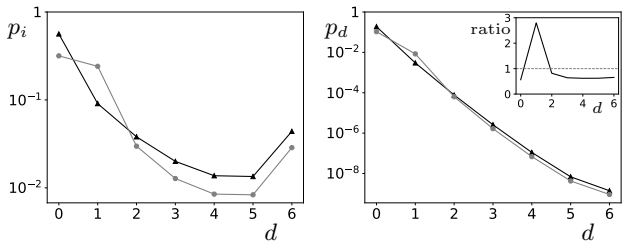

\textit{Maximization of the overlap with multiple spins}. We will now go beyond the standard result of \cite{CHVA} to highlight the flexibility of our approach. Consider $w_a=1$ for $\hat{S}^z_{a}$ on three adjacent sites, $a=\{j-1,j,j+1\}$, so the $Q$-matrix reads
\beq\label{3sites}
Q_{mn}=\sum_{i=j-1}^{j+1}\langle n|\hat{S}_i^z|n\rangle\langle m|\hat{S}_i^z|m\rangle.
\eeq
As in the previous paragraph, the optimization problem is solved by finding the top eigenvector of $Q$, but now the integral of motion has more support on three adjacent sites, and reduced overlap outside this region, see Fig.~\ref{fig:od} for comparison. In order to visualise a LIOM $\hat V$ we first remove its trace, $\tilde V = \hat V-{2^{-L}}\Tr[\hat V]\,\mathbb I$, and then expand the traceless operator $\tilde V$ in the Pauli-string basis  $\{P_\alpha\}$, where $P_\alpha=\bigotimes_{i=1}^{L}\sigma_i^{\mu_i}$ with 
$\mu_i\in\{0,x,y,z\}$ and $\sigma_i^{0}=I$. The weight of each string is given by $p_\alpha = 2^{-L}(\Tr[\tilde VP_\alpha])^{2}$, and their sum equals one. A string $P_\alpha$ has support on the sites where $\sigma_i^{\mu_i}\neq I$.  We assign the weight $p_\alpha$ to the site in this support that lies farthest from the center of the chain (if two sites are equally distant we split $p_\alpha$ evenly between them). This prescription produces a site-resolved probability distribution $p_i$ for a given integral of motion. Averaging over all operators at a fixed distance $d$ from the central site at $i_0$ yields $p_d = \frac{1}{N(d)}\sum_{i: |i-i_0|=d} p_i$, where $N(d)$ is the number of operators at distance $d$ from $i_0$. As the number of Pauli strings grows exponentially with their length, the total probability $p_i$ decays slower than the averaged per-operator quantity $p_d$. (We provide some extra numerical details of LIOM localization in terms of $p_d$ in the supplemental material.)

\textit{Maximization of the overlap with a single spin under eigenvalue constraints}.  
Now, our task is to construct a LIOM $\hat V$ that (i) has maximum overlap with a given spin $\hat S^{z}_{j}$, and (ii) possesses the same binary spectrum as a single spin-$\tfrac12$ degree of freedom $v_{n}\in\{-1,+1\}$.  After inserting the single-spin $Q$-matrix \eqref{Qsing} into the objective function \eqref{Rv} and defining $c^{\,j}_{n}\equiv \langle n|\hat S^{z}_{j}|n\rangle$, we reduce the problem to the maximization of a rank-1 quadratic form:
\beq\label{rank1}
R^{j}(\mathbf v)=(\mathbf c^{\,j}\cdot\mathbf v)^{2},\ \ \ \ v_{n}\in\{-1,+1\}.
\eeq
The solution of this problem is given by the vector $v_{n}= \operatorname{sign}(c^{\,j}_{n})$, up to an overall sign, or in other words, by the vertex of the hypercube closest to the direction $\mathbf c^{\,j}$. In order to have equal number of $\,\pm1$ eigenvalues, one has to impose an additional constraint $\sum_{n}v_{n}=0$.  In practice, we sort the values of $c^{\,j}_{n}$ by their magnitude, followed by the assignment of $v_{n}=-1$ to the lower half of the list, and $v_{n}=+1$ to the upper half. This construction coincides with the one employed in Ref.~\cite{PLYWC}, where it was shown to maximize the overlap between the LIOM and the target single-site spin operator.

\textit{Maximization of the overlap with multiple adjacent spins under eigenvalue constraints}.  
We now consider a target of single-spin operators from a block $\mathcal A=\{j,j+1,\dots,j+M-1\}$ of $M$ adjacent sites. The objective function is a quadratic form whose rank generically equals $M$:
\begin{equation}
R^{\mathcal A}(\mathbf v)=\sum_{s\in \mathcal A}(\mathbf c^{s}\!\cdot\!\mathbf v)^{2},
\qquad
v_{n}\in\{-1,+1\},
\label{eq:multi_spin_obj}
\end{equation}
so that the optimization problem reads $\max_{\mathbf v\in\{\pm1\}^{D}}R^{\mathcal A}$, subject to an optional constraint $\sum_{n}v_{n}=0$.
This is a QUBO problem defined on the corners of a hypercube. For the problem of optimization of the overlap with only a small subset of sites, the scaling of complexity is polynomial $D^{\mathcal{O}(M)}$. The operators $\hat V$ obtained from this optimization overlap predominantly with the chosen set of adjacent spins, while retaining the binary eigenvalue spectrum $\pm1$. This construction constitutes the most elaborate scheme presented in this work.

By identifying $Q_{mn}$ with $-J_{mn}$ (coupling constants) and $v_n$ with $\sigma^{z}_n$ (classical spins), our optimization problem is equivalent to finding the ground state of a classical spin Hamiltonian, $H=-{\sum}_{ij}\,J_{ij}\sigma^{z}_i\sigma^{z}_j$.
In our case, the coupling matrix $J_{ij}$ is low-rank, but contains long-range interactions with positive and negative signs, reminiscent of spin-glass models, such as e.g.~the Sherrington-Kirkpatrick model \cite{SKM}. It is also similar to the generalized Hopfield neural network \cite{HopN,HopSG}, where the interaction matrices are similarly of low rank. Finding ground states of spin-glass models is generically an NP-hard problem, which becomes polynomial in complexity when the coupling matrix has a low rank.

\textit{Optimization for rank-2 couplings}. In the case of overlap with two sites $M=2$, the `angle-sorting' algorithm \cite{angle} solves the rank-2 problem by breaking it down into multiple rank-1 optimizations, which can be addressed with the methods outlined above. In the presence of the constraint $v_n=\pm1$, the reward function reads $R(\mathbf{v})=(\mathbf{c}^j\cdot\mathbf{v})^2 + (\mathbf{c}^{j+1}\cdot\mathbf{v})^2$. We can turn this into a rank-1 problem by introducing an auxiliary angle $\theta$, and a reward function $\mathcal{R}(\mathbf{v},\theta)=\big[(\mathbf{c}^j\cdot\mathbf{v})\cos{\theta} + (\mathbf{c}^{j+1}\cdot\mathbf{v})\sin{\theta}\big]^2$.
One can see that, for any given $\mathbf{v}$, the maximum of $\mathcal{R}(\mathbf{v},\theta)$ with respect to $\theta$ is obtained when the direction of $\theta$ aligns with the 2d vector $(\mathbf{c}^j\cdot\mathbf{v},\hspace{0.1cm} \mathbf{c}^{j+1}\cdot\mathbf{v})$. At this maximum, one recovers the solution as $\max_\theta\mathcal{R}(\mathbf{v},\theta)=R(\mathbf{v})$, and the optimization problem reads
\beq\label{rgen}
\max_\mathbf{v} R(\mathbf{v})=\max_{\mathbf{v},\theta}\mathcal{R}(\mathbf{v},\theta).
\eeq
Expressing the auxilary reward function as $\mathcal{R}(\mathbf{v},\theta)=[(\mathbf{c}^j\cos{\theta} + \mathbf{c}^{j+1}\sin{\theta})\cdot\mathbf{v}]^2$ reduces it to a rank-1 form, as in \eqref{rank1}, and one can now readily maximize it over $\mathbf{v}$ for any fixed $\theta$. The solution is given by $v_n(\theta)=\text{sgn}(c_n^j\cos{\theta} + c_n^{j+1}\sin{\theta})$, which is piecewise constant in $\theta$, since it only changes at the angles $\theta_n$ where one of the arguments vanishes:
\beq\label{eq_cn}
c_n^j\cos{\theta_n} + c_n^{j+1}\sin{\theta_n}=0,\quad n=1\ldots D.
\eeq
Owing to the symmetry $\mathcal{R}(\mathbf{v},-\theta)=\mathcal{R}(\mathbf{v},\theta)$, the angles are restricted to $0\leq\theta<\pi$ and (\ref{eq_cn}) has a single solution for each $n$. Given this set of angles $\theta_n$, we can sort them on the semicircle, which results in separation into $D+1$ sections. In order to maximize \eqref{rgen}, we pick an angle from each section, assign the optimal $v_n$ given above, and evaluate the reward function. Since the assignment of this $v_n$ is constant inside each sector, the solution is given by the $v_n$ in the sector where the reward function is the largest.
Using the same strategy, we can also introduce the further constraint of balanced eigenvalues to the two-site optimization. The procedure remains the same, but now in order to maximize \eqref{rgen} for a given angle $\theta$ we must order the values $c_n^j\cos{\theta} + c_n^{j+1}\sin{\theta}$ for all $n$, and assign $v_n=-1$ to the smallest half and $v_n=+1$ to the rest. This assignment is still piecewise-constant in $\theta$, but now it may change whenever the angle $\theta=\theta_{mn}$ is such that a pair of values $(m,n)$ gives equal components:
\[
c_n^j\cos{\theta_{mn}} + c_n^{j+1}\sin{\theta_{mn}}=c_m^j\cos{\theta_{mn}} + c_m^{j+1}\sin{\theta_{mn}}.
\]
Once again, the required integral of motion can be constructed by finding all solutions, arranging them around the semicircle, computing $\mathcal{R}(\mathbf{v},\theta)$ for each angle and the corresponding $\mathbf{v}$, and picking the $\mathbf{v}$ that yields the global maximum. In principle, this algorithm can be extended to integrals of motion localized on more than two sites \cite{angle}. However, rank 3 already appears intractable due to excessive computation times even for chains of moderate length, at least without further optimization.

\begin{figure}[t]
\centering
\mbox{\subfigure{\includegraphics[width=0.23\textwidth]{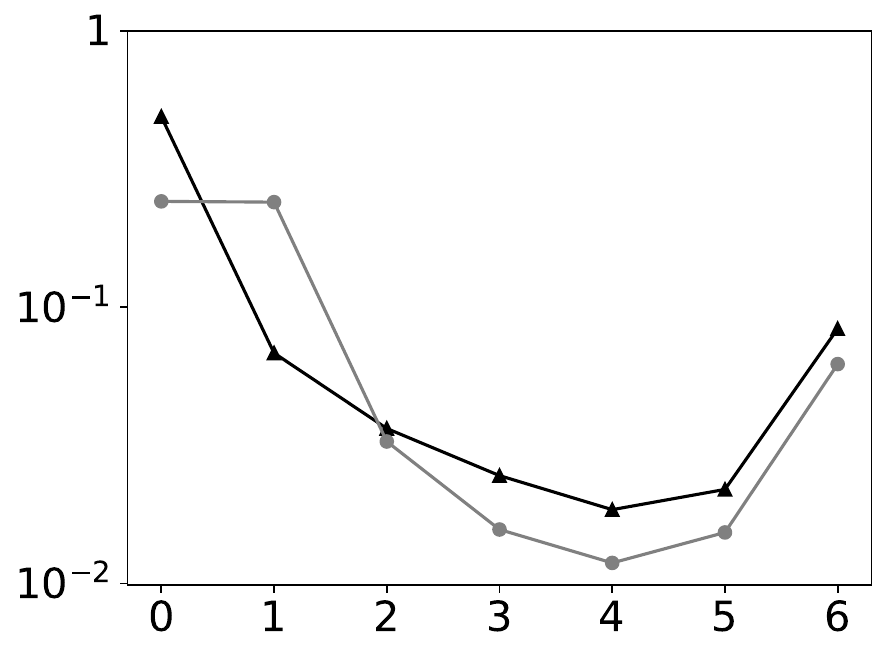}}\,
 \subfigure{\includegraphics[width=0.23\textwidth]{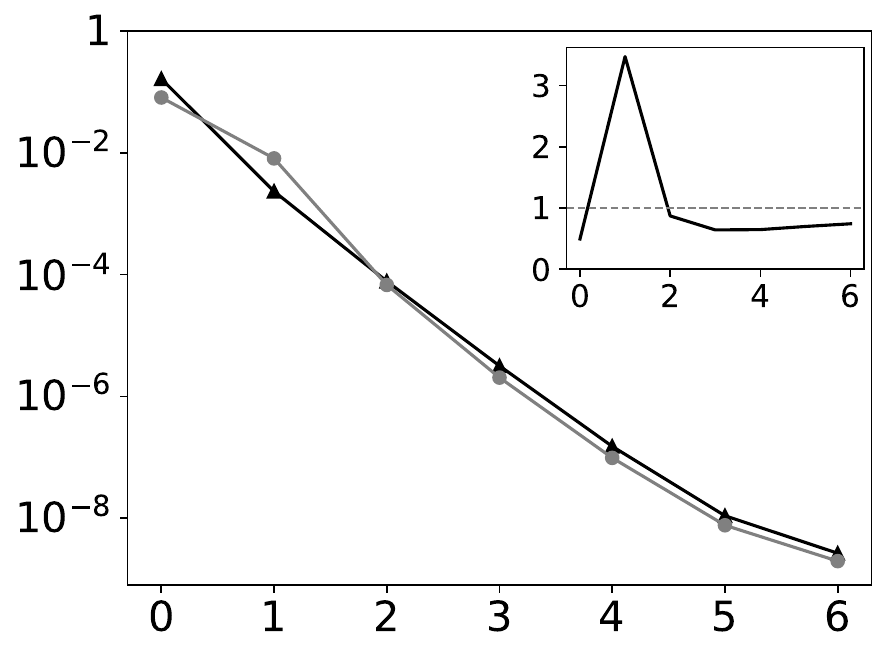}}}
\begin{picture}(0,0)
\put(-140,-5){$d$}
\put(-237,73){$p_i$}
\put(-20,-5){$d$}
\put(-117,73){$p_d$}
\put(-62,69){$\mathrm{\scriptstyle ratio}$}
\put(-16,43){$\scriptstyle d$}
\end{picture}
\caption{Same as in Fig.~\ref{fig:od}, but in the presence of additional constraints $v_n=\pm1$ and $\sum_n v_n=0$.}
\label{fig:odbal}
\end{figure}
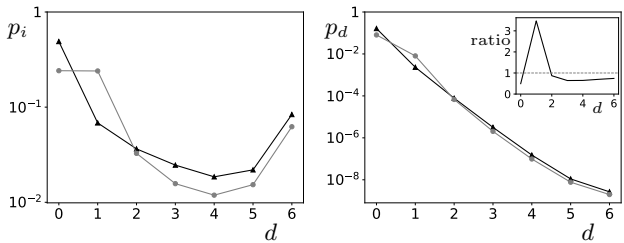

\textit{Higher-rank cases.} In addition to the optimization methods mentioned above, this class of problems can be studied with commercial optimization software such as Gurobi \cite{gurobi}. In broad terms, this state-of-the-art solver uses a branch-and-bound approach along with continuous relaxation to solve mixed integer problems such as the ones relevant to our construction, for a review on various optimization approaches, see \cite{Kochenberger}. To give an example, one could fix some number of $v_n=\pm1$ and solve the smaller and unconstrained problem for the rest of the $v_n$, which are allowed to take non-integer values. If the reward for this solution is worse than the current best known solution, this branch can be pruned. The solver combines this approach with other more advanced methods to significantly reduce computational time. It has been used to manage similar integer optimization problems in a wide range of applications such as manufacturing logistics, sports scheduling, and industrial process scheduling \cite{logistics,sports,scheduling}. For our purposes, this solver makes it possible to run the optimization routine and construct LIOMs with constrained eigenvalues centered on more than two adjacent sites. In Fig.~\ref{fig:odbal}, we compare the results of this constrained optimization on the central site and on three adjacent sites, which shows that the latter has reduced support away from the central site.

A comment on the computational complexity of the algorithms employed above: for the unconstrained optimization, both on one and three sites, the scaling is $O(D^3)$ (with $D\equiv2^L$) due to the matrix operations involved. The total computational time is dominated by the diagonalization of the Hamiltonian, which is a required step of the construction. The constrained maximization problem on a single site has the same scaling, again due to the matrix multiplications involved, but now its contribution to the total time exceeds that of the initial diagonalization step. Optimization on three sites involves solving the mixed integer problem with Gurobi, for which there are no published runtimes. Empirically, it exhibits high variability across instances, but generically dominates all other parts of the computations (over the ranges we tested). We have made the numerical scripts public \cite{zenodo}.\vspace{1mm}

To summarize, we have presented a general method for constructing LIOMs with maximized support over a chosen set of local operators. Without extra constraints on the eigenvalues, the problem reduces to a simple maximization of a quadratic form. Constraining the eigenvalues to specific discrete choices results in more sophisticated maximization problems, which are reminiscent of spin-glass physics, and we have outlined and implemented practical methods for their solution. By employing the prototypical Heisenberg model, we have constructed a variety of LIOMs, which we optimized according to diverse locality criteria. Simple realizations of this protocol allow one to systematically recover a number of LIOM constructions previously discussed in the literature. The flexibility of our `designer LIOM' scheme leads to a variety of more sophisticated constructions. For instance, allowing for greater support of LIOMs over a few sites adjacent to the central site lets one suppress the tails of their support over more remote regions. We have also shown how the related constrained problems arising for LIOMs with spin-like eigenvalues can be solved with either simple intuitive methods or commercial state-of-the-art optimization software.

An important application of LIOMs is that they can be utilized to study dynamics in the MBL phase in useful ways \cite{SPA}. If we construct these integrals of motion in the form (\ref{Hvn}), with each one localized on (or around) a different site, generically we can use them to expand the Hamiltonian in terms of these operators:
\beq\label{decomp}
\hat H=\sum_i c_i\hat\tau_i+\sum_{i>j}c_{ij}\hat\tau_i\hat\tau_j+\sum_{i>j>k}c_{ijk}\hat\tau_i\hat\tau_j\hat\tau_k+\dots.
\eeq
Indeed, for generic sets of eigenvalues $v_n$, taking products of such LIOMs defined by (\ref{Hvn}) will produce new linearly independent sets of $v_n$, eventually exhausting all independent directions in the space of conserved operators. Therefore, the expansion shown in (\ref{decomp}) is just an expansion of the spin-Hamiltonian in this complete basis. 

There are extra subtleties in the expansion (\ref{decomp}) if the $\tau_i$ are constrained to have the same eigenvalues as the spin operators, in which case the decomposition \eqref{decomp} expresses the Hamiltonian in terms of a set of $L$ `pseudospins.' Such a decomposition is not always possible, in contrast to LIOMs with unconstrained eigenvalues, because this set of constrained $\hat\tau_i$ and their products do not necessarily form a complete basis in the space of conserved operators. Indeed, if $v_n$ are only allowed to take values $\pm 1$, evaluating their products does not necessarily generate new sets of $v_n$ linearly independent of the previous ones. Optimizing all $L$ LIOMs at once to get a complete set suitable for the expansion (\ref{decomp}) while including the eigenvalue constraint is an intriguing problem that we leave for future work. A possible (but likely not optimal) construction for the case of single-spin LIOMs satisfying $v_{n}\in\{-1,+1\}$ and $\sum_{n}v_{n}=0$ was proposed in \cite{PLYWC}. Extra investigation is required to extend this to more general cases. However, the need for constraining the eigenvalues depends on the practical applications one has in mind. For example, in studies of dynamics in MBL, a simple strategy could be to drop the eigenvalue constraint, in which case the construction of a complete set of LIOMs on all sites can be done independently site-by-site.

The expansion (\ref{decomp}) in principle applies to any Hamiltonian and a sufficiently large set of its conservation laws. However, for local Hamiltonians that possess LIOMs, our construction becomes considerably simpler and more useful, since any term containing a pair of distant sites should be strongly suppressed. This creates room for exploring dynamics of longer chains, since, in the localized phase, it should be possible to construct each $\hat\tau_i$ with a required precision by only considering a small segment of the chain around the central site. The ability to suppress the tails of $\hat\tau_i$ away from the central site that arises naturally from our framework is a very welcome addition from this perspective as---together with the tails---one represses the sensitivity of LIOMs to finite-size truncations.\vspace{3mm}

\begin{acknowledgments} {\it Acknowledgments}.
We thank Wojciech De Roeck for discussions. Work at VUB was supported by FWO-Vlaanderen projects G012222N and G0A2226N and by the VUB Research Council through the Strategic Research Program High-Energy Physics. O.E. is supported by the  C2F program at Chulalongkorn University and by NSRF via grant number B41G680029. D.K. acknowledges support from Labex MME-DII grant ANR11-LBX-0023, and funding under The Paris Seine Initiative Emergence programme 2019. G.P. is supported by a PhD fellowship from FWO. The resources and services used in this work were provided by the VSC (Flemish Supercomputer Center), funded by the Research Foundation Flanders (FWO) and the Flemish Government.
\end{acknowledgments}

\onecolumngrid

\onecolumngrid

\newpage

\centerline{\large\bf Supplemental material}\vspace{1cm}

\twocolumngrid

\subsection*{LIOMs in the Anderson model}

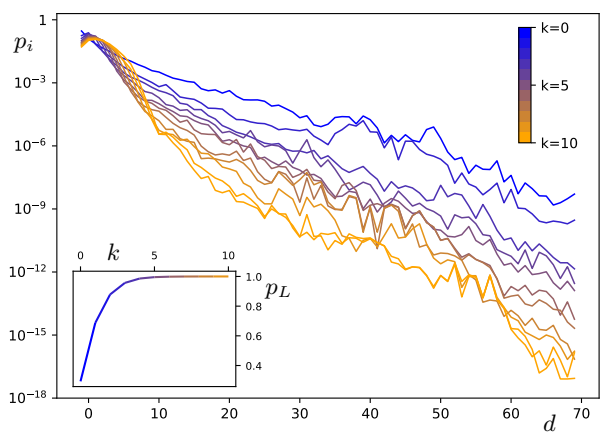
\begin{figure}[b]
\centering
\includegraphics[width=0.46\textwidth]{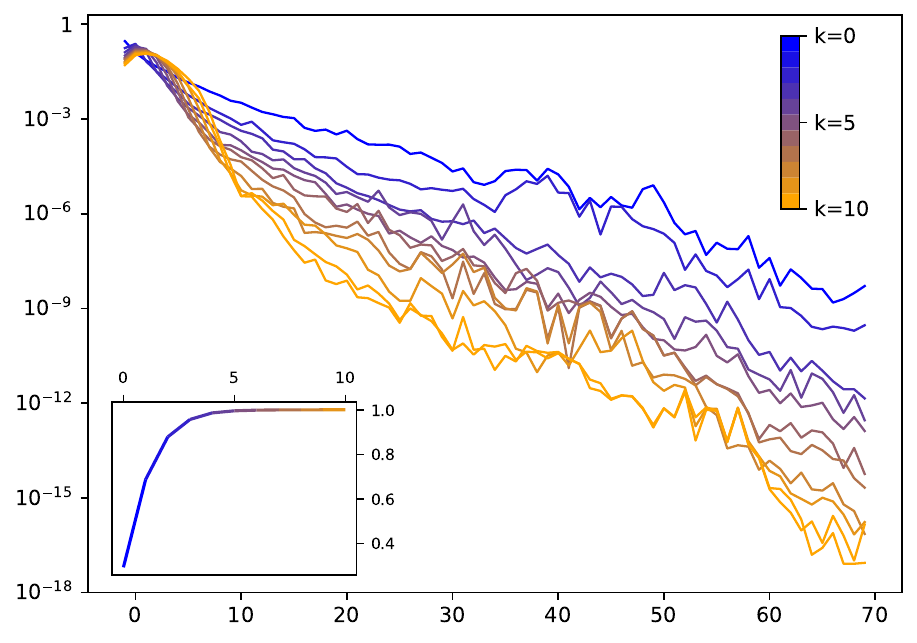}
\put(-25,0){$d$}
\put(-225,145){{$p_i$}}
\put(-190,65){$k$}
\put(-130,52){$p_L$}
\caption{Support probability (log scale) of the integrals of motion in the Anderson chain \eqref{andham} with 1000 sites and disorder $W=5$, as a function of distance from the central site $i=500$, averaged over 1000 realizations. The curves correspond to integrals of motion with maximized support on $k$ sites around the center, with varying $k$. \textit{Inset:} Total probability of the local set of sites $(i-k,\dots i+k)$ for each integral of motion.}
\label{fig:ander}
\end{figure}

We consider a tight binding chain of length $L$ with periodic boundary conditions:
\beq\label{andham}
\hat{H}_d =-t\sum_{\langle ij\rangle}|i\rangle\langle j|+\sum_i \epsilon_i |i\rangle\langle i|. 
\eeq
The diagonal terms $\epsilon_i$ describe onsite disorder which leads to localization of the conservation laws. The disorder is drawn uniformly from $[-W/2, W/2]$. In what follows, we set the tunneling strength $t=1$. Again, we denote the eigenstates of $H$ by $|n\rangle$. The site basis is denoted by $|i\rangle$. To start, we choose to localize on a single site, so we assign weight 1 only to the projector $\displaystyle |i\rangle\langle i|$. The resulting $Q$-matrix is simply $Q_{mn}=\langle n|i\rangle\langle i|n\rangle\langle m|i\rangle\langle i|m\rangle$. Solving for the eigenvector corresponding to the largest eigenvalue leads to the following (unnormalized) integral of motion:
\beq
    \mathcal{I}_i=\sum_n |\langle i|n\rangle|^2 |n\rangle\langle n|.
\eeq
We can repeat the computations with a larger set of favored operators: for example, the set of operators within a distance $k$ of site $i$: $\{|j\rangle\langle l| \text{ with }j,l\in (i-k,\dots,i,\dots,i+k)\}$, where $k=0$ reproduces the previous computation. This set should be made Hermitian and orthogonal before constructing the $Q$-matrix, as in \eqref{apbasis}. Clearly, assigning nonzero weight to more sites allows the support range to grow, while the tails far away from the center become more suppressed. Similarly to the main text, to display the locality of an integral of motion $V$, we first compute probabilities over a complete basis $\mathcal{B}_L=\{P_\alpha\}$ given by
\beq
p_\alpha = \Tr[V P_\alpha]^2,
\eeq
with
\begin{align}
\label{apbasis}\mathcal{B}_L = \big\{
|i\rangle\langle i|,
\tfrac{1}{\sqrt{2}}(|i\rangle\langle j|& + |j\rangle\langle i|),
\tfrac{i}{\sqrt{2}}(|i\rangle\langle j| - |j\rangle\langle i|)
\nonumber\\
&\text{for all } i,j \text{ with } j>i
\big\}.
\end{align}
We then distribute each of these to its respective site $i$ or $j$, whichever is most distant from the center of the chain. For weight 1 assigned to a bigger range of central sites, the results displayed in Fig.~\ref{fig:ander} show increased support on these sites, accompanied by stronger suppression of the tails.\vspace{5mm}

\subsection*{Localization length and suppression\\of the tails of LIOMs}

It is useful to quantify more precisely how the LIOMs we construct are localized, and how this localization differs between the LIOMs localized on one site and three sites. To do this, for each disorder realization, we take the average probability $p_d$ over all operators at distance $d$ from the central site, as introduced in the main text. We then parametrize the asymptotic tails as 
\beq\label{pfit}
p_d=Ae^{-d/\xi}
\eeq
at sufficiently large $d$ (evidently, the range of $d$-values accessible to our small-chain simulations is limited, and different fitting procedures can be proposed, see below). In (\ref{pfit}), $\xi$ is the localization length and $A$ is the overall normalization of the tail. 
\begin{table*}[t]
\caption{Results of the exponential fit (\ref{pfit}), with $(\xi_1,A_1)$ referring to 1-site LIOMs and $(\xi_3,A_3)$ to the 3-site LIOMs. Each result is given as the mean $\pm$ standard error over $1000$ disorder realizations. We report results when fitting (\ref{pfit}) over two distance regions: all available distances, with the three-site construction starting from the edge of the LIOM $d=1$, and only the larger distances $d=4,5,6$.}
\label{tab:fit}
\begin{ruledtabular}
\begin{tabular}{ccccc}
 & Unconstrained & Constrained & Unconstrained ($d=4,5,6$) & Constrained ($d=4,5,6$) \\
$\xi_1$ & $0.3139 \pm 0.0005$ & $0.3298 \pm 0.0006$ & $0.485 \pm 0.003$ & $0.545 \pm 0.005$ \\
$\xi_3$ & $0.3122 \pm 0.0006$ & $0.3332 \pm 0.0008$ & $0.503 \pm 0.003$ & $0.590 \pm 0.006$ \\
$A_1$ & $0.0640 \pm 0.0007$ & $0.0472 \pm 0.0006$ & $0.0020 \pm 0.0002$ & $0.0021 \pm 0.0003$ \\
$A_3$ & $0.0533 \pm 0.0009$ & $0.0402 \pm 0.0009$ & $0.0010 \pm 0.0001$ & $0.0011 \pm 0.0002$ \\
\end{tabular}
\end{ruledtabular}
\end{table*}
As the results of Table~\ref{tab:fit} show, the average localization length changes only slightly between the two constructions. It is natural to assume that in the limit of long chains, the effect of different optimization choices (one-site vs.\ three-site) on the localization length is negligible, and it should  physically be governed by the localization length of the energy eigenstates. The dominant effect on the tail probability suppression comes instead from the difference in the tail normalization $A$ between the one-site and three-site LIOMs ($A_3$ is reliably smaller than $A_1$ for all fitting specifications, even though we can only probe the tail over a very limited range of $d$). This provides an effective mechanism for tail suppression in the three-site LIOMs, even in the absence of substantial reduction of the localization length.


\begin{thebibliography}{99}\vspace{-2mm}
\twocolumngrid

\bibitem{MBL1}I.~V.~Gornyi, A.~D.~Mirlin and D.~G.~Polyakov, {\it Interacting electrons in disordered wires: Anderson localization and low-$T$ transport,}
\doi{Phys.\ Rev.\ Lett. {\bf 95} (2005) 206603}{10.1103/PhysRevLett.95.206603}, \arXiv{cond-mat/0506411}.

\bibitem{MBL2}D.~M.~Basko, I.~L.~Aleiner and B.~L.~Altshuler,\\ {\it Metal-insulator transition in a weakly interacting many-electron system with localized single-particle states,}\\ \doi{Ann.\ Phys. {\bf 321} (2006) 1126}{10.1016/j.aop.2005.11.014}, \arXiv{cond-mat/0506617}.

\bibitem{MBL3}
D.~A.~Huse, R.~Nandkishore and V.~Oganesyan,
{\it Phenomenology of fully many-body-localized systems,}
\doi{Phys.\ Rev.\ B \textbf{90} (2014) 174202}{10.1103/PhysRevB.90.174202},
\arXiv{1408.4297} [cond-mat.stat-mech].

\bibitem{review}
P.~Sierant, M.~Lewenstein, A.~Scardicchio, L.~Vidmar and J.~Zakrzewski,
{\it Many-body localization in the age of classical computing,}
\doi{Rept. Prog. Phys. \textbf{88} (2025) 026502}{10.1088/1361-6633/ad9756},
\arXiv{2403.07111} [cond-mat.dis-nn].

\bibitem{anderson}
P.~W.~Anderson,
{\it Absence of diffusion in certain random lattices,}
\doi{ Phys. Rev. \textbf{109} (1958) 1492}{10.1103/PhysRev.109.1492}.

\bibitem{Bloch}
M.~Schreiber, S.~S.~Hodgman, P.~Bordia, H.~P.~L{\"u}schen, M.~H.~Fischer, R.~Vosk, E.~Altman, U.~Schneider and I.~Bloch,
{\it Observation of many-body localization of interacting fermions in a quasirandom optical lattice,}
\doi{Science \textbf{349} (2015) aaa7432}{10.1126/science.aaa7432},
\arXiv{1501.05661} [cond-mat.quant-gas].

\bibitem{Greiner}
M.~Rispoli, A.~Lukin, R.~Schittko, S.~Kim, M.~E.~Tai, J.~L{\'e}onard and M.~Greiner,
{\it Quantum critical behaviour at the many-body localization transition,}
\doi{Nature \textbf{573} (2019) 385}{10.1038/s41586-019-1527-2},
\arXiv{1812.06959} [cond-mat.quant-gas].

\bibitem{Roushan}
P.~Roushan \textit{et al.},
{\it Spectroscopic signatures of localization with interacting photons in superconducting qubits,}
\doi{Science \textbf{358} (2017) 1175}{10.1126/science.aao1401},
\arXiv{1709.07108} [quant-ph].

\bibitem{Shtanko}
O.~Shtanko, D.~S.~Wang, H.~Zhang, N.~Harle, A.~Seif, R.~Movassagh and Z.~Minev,
{\it Uncovering local integrability in quantum many-body dynamics,}
\doi{Nature Comm. \textbf{16} (2025) 2552}{10.1038/s41467-025-57623-x},
\arXiv{2307.07552} [quant-ph].

\bibitem{SPA}M.~Serbyn, Z.~Papi\'c and D.~A.~Abanin, {\it Local conservation laws and the structure of the many-body localized states,}
\doi{Phys.\ Rev.\ Lett. {\bf 111} (2013) 127201}{10.1103/PhysRevLett.111.127201},\\\arXiv{1305.5554} [cond-mat.dis-nn].

\bibitem{RosMuellerScardicchio}V.~Ros, M.~M\"uller and A.~Scardicchio, {\it Integrals of motion in the many-body localized phase,} \doi{Nucl.\ Phys.\ B
{\bf 891} (2015) 420}{10.1016/j.nuclphysb.2014.12.014}, \arXiv{1406.2175} [cond-mat.dis-nn].

\bibitem{CHVA}A.~Chandran, I.~H.~Kim, G.~Vidal and D.~A.~Abanin, {\it Constructing local integrals of motion in the many-body localized phase,}
\doi{Phys.\ Rev.\ B {\bf 91} (2015) 085425}{10.1103/PhysRevB.91.085425}, \arXiv{1407.8480} [cond-mat.dis-nn].

\bibitem{RO}L.~Rademaker and M.~Ortu\~no, {\it Explicit local integrals of motion for the many-body localized state,}
\doi{Phys.\ Rev.\ Lett. {\bf 116} (2016) 010404}{10.1103/PhysRevLett.116.010404}, \arXiv{1507.07276} [cond-mat.str-el].

\bibitem{OAVP}T.~E.~O'Brien, D.~A.~Abanin, G.~Vidal and Z.~Papi\'c, {\it Explicit construction of local conserved operators in disordered many-body systems,}
\doi{Phys.\ Rev.\ B {\bf 94} (2016) 144208}{10.1103/PhysRevB.94.144208}, \arXiv{1608.03296} [cond-mat.dis-nn].

\bibitem{PLYWC}P.~Peng, Z.~Li, H.~Yan, K.~X.~Wei and P.~Cappellaro, {\it Comparing many-body localization lengths via non-perturbative construction of local integrals of motion,} \doi{Phys.\ Rev.\ B {\bf 100} (2019) 214203}{10.1103/PhysRevB.100.214203}, \arXiv{1901.00034} [cond-mat.dis-nn].

\bibitem{AAS}S.~Adami, M.~Amini and M.~Soltani, {\it Structural properties of local integrals of motion across the many-body localization transition via a fast and efficient method for their construction,} \doi{Phys.\ Rev.\ B {\bf 106} (2022) 054202}{10.1103/PhysRevB.106.054202}, \arXiv{2204.08835} [cond-mat.dis-nn].

\bibitem{recent1}
J.~Paw\l owski, J.~Herbrych and M.~Mierzejewski,
{\it Finding local integrals of motion in quantum lattice models in the thermodynamic limit,}
\doi{Phys. Rev. B \textbf{112} (2025) 155130}{10.1103/brst-wp36},\\
\arXiv{2505.05882} [cond-mat.str-el].

\bibitem{recent2}
J.~Paw{\l}owski, P.~{\L}yd{\.z}ba and M.~Mierzejewski,\\
{\it Local integrals of motion encoded in a few eigenstates,}
\arXiv{2603.01859} [cond-mat.str-el].

\bibitem{recent3}
T.~H.~Yang, S.~Gopalakrishnan and D.~A.~Abanin,\\
{\it Simple slow operators and quantum thermalization,}
\arXiv{2604.13172} [quant-ph].

\bibitem{SKM}
D.~Sherrington and S.~Kirkpatrick,
{\it Solvable model of a spin-glass,} \doi{Phys.\ Rev.\ Lett. {\bf 35} (1975) 1792}{10.1103/PhysRevLett.35.1792}.

\bibitem{HopN}
J.~J.~Hopfield,
{\it Neural networks and physical systems\\ with emergent collective computational abilities,}\\\doi{PNAS {\bf 79} (1982) 2554}{10.1073/pnas.79.8.2554}.

\bibitem{HopSG}
D.~J.~Amit, H.~Gutfreund and H.~Sompolinsky,\\
{\it Spin-glass models of neural networks,}\\
\doi{Phys.\ Rev.\ A {\bf 32} (1985) 1007}{10.1103/PhysRevA.32.1007}.

\bibitem{bound}
B.~Craps, M.~De Clerck, O.~Evnin, P.~Hacker and M.~Pavlov,
{\it Bounds on quantum evolution complexity via lattice cryptography,} \doi{SciPost Phys. \textbf{13} (2022) 090}{10.21468/SciPostPhys.13.4.090},
 \arXiv{2202.13924} [quant-ph].

\bibitem{complint}
B.~Craps, M.~De Clerck, O.~Evnin and P.~Hacker,
{\it Integrability and complexity in quantum spin chains,} \doi{SciPost Phys. \textbf{16} (2024) 041}{10.21468/SciPostPhys.16.2.041},
\arXiv{2305.00037} [quant-ph].

\bibitem{Anderson_Ising}Y.~Fu and P.~W.~Anderson, {\it Application of statistical mechanics to NP-complete problems in combinatorial optimisation}, 
\doi{J.\ Phys.\ A \textbf{19} (1986) 1605}{10.1088/0305-4470/19/9/033}.

\bibitem{Pardalos}P.~M.~Pardalos and S.~Jha, {\it Complexity of uniqueness\\ and local search in quadratic 0–1 programming},\\ 
\doi{Oper.\ Res.\ Lett. \textbf{11} (1992) 119}{10.1016/0167-6377(92)90043-3}.

\bibitem{Kochenberger}G.~Kochenberger, J.-K.~Hao, F.~Glover, M.~Lewis, Zh.~L\"u, H.~Wang and Y.~Wang,
 {\it The unconstrained\\ binary quadratic programming problem: a survey},\\ 
\doi{J.\ Comb.\ Optim. \textbf{28} (2014) 58} {10.1007/s10878-014-9734-0}.

\bibitem{transition}
A.~Pal and D.~A.~Huse,
{\it The many-body localization\\ phase transition,} \doi{Phys. Rev. B \textbf{82} (2010) 174411}{10.1103/PhysRevB.82.174411},\\
\arXiv{1010.1992} [cond-mat.dis-nn].

\bibitem{quspin}P.~Weinberg and M.~Bukov, {\it QuSpin: a Python package for dynamics and exact diagonalisation of quantum many body systems. Part I: spin chains,}
\doi{SciPost Phys. {\bf 2} (2017) 003}{10.21468/SciPostPhys.2.1.003}, \arXiv{1610.03042} [physics.comp-ph].

\bibitem{angle}
G.~N.~Karystinos and A.~P.~Liavas,
{\it Efficient computation of the binary vector that maximizes a rank-deficient quadratic form,}
\doi{IEEE Trans.\ Inf.\ Th. {\bf 56} (2010) 3581}{10.1109/TIT.2010.2048450}.

\bibitem{gurobi}
Gurobi Optimization, LLC, {\it Gurobi Optimizer Reference Manual,} \href{www.gurobi.com}{www.gurobi.com} (2025).

\bibitem{logistics}
R.~Baller, P.~Fontaine, S.~Minner and Z.~Lai,
{\it Optimizing automotive inbound logistics: A mixed-integer linear programming approach,}
\doi{Transp.\ Res.\ E {\bf 163} (2022) 102734}{10.1016/j.tre.2022.102734}.

\bibitem{sports}
R.~Lambers, L.~Rothuizen and F.~C.~R.~Spieksma,\\
{\it How to schedule the Volleyball Nations League,}\\
\doi{J.\ Sports\ Anal. {\bf 9} (2023) 157}{10.3233/JSA-220626}.

\bibitem{scheduling}
T.~J.~Ikonen, K.~Heljanko and I.~Harjunkoski,
{\it Surrogate-based optimization of a periodic rescheduling algorithm,}
\doi{AIChE J.\ {\bf 68} (2022) e17656}{10.1002/aic.17656}.

\bibitem{zenodo}
\doi{Zenodo 17877431 (2025)}{10.5281/zenodo.17877431}.

\end{thebibliography}
\end{document}